\documentclass[preprint,superscriptaddress,nofootinbib]{revtex4}
\usepackage{amsmath}
\usepackage{amssymb}
\usepackage{graphicx}
\usepackage{mathrsfs}
\usepackage{color}
\usepackage{multirow}

\newcommand{\beq}{\begin{equation}}
\newcommand{\eeq}{\end{equation}}
\newcommand{\bea}{\begin{eqnarray}}
\newcommand{\eea}{\end{eqnarray}}

\newcommand{\shat}{\hat{s}}

\allowdisplaybreaks[1]

\begin{document}

\title{Searching for charged Higgs boson in polarized top quark}

\author{Qing-Hong Cao }
\email{qinghongcao@pku.edu.cn}
\affiliation{Institute of Theoretical Physics and State Key Laboratory of
Nuclear Physics and Technology, Peking University, Beijing 100871,
China }

\author{Xia Wan}
\email{xia.wan@pku.edu.cn}
\affiliation{Institute of Theoretical Physics $\&$ State Key Laboratory of
Nuclear Physics and Technology, Peking University, Beijing 100871,
China }

\author{Xiao-ping Wang}
\email{hcwangxiaoping@pku.edu.cn}
\affiliation{Institute of Theoretical Physics $\&$ State Key Laboratory of
Nuclear Physics and Technology, Peking University, Beijing 100871,
China }

\author{Shou-hua Zhu}
\email{shzhu@pku.edu.cn}
\affiliation{Institute of Theoretical Physics $\&$ State Key Laboratory of
Nuclear Physics and Technology, Peking University, Beijing 100871,
China }
\affiliation{Center for High Energy Physics, Peking University, Beijing 100871, China}

\date{\today}

\begin{abstract}
The charged Higgs boson is quite common in many new physics models. In this study we examine the potential of observing a heavy charged Higgs boson in its decay mode of top quark and bottom quark in the type-II two-Higgs-doublet model. In this model, the chirality structure of the coupling of a charged Higgs boson to the top and bottom quark is very sensitive to the value of $\tan\beta$. As the polarization of the top quark can be measured experimentally from the top quark decay products, one could make use of the top quark polarization to determine the value of $\tan\beta$. We perform a detailed analysis of measuring top quark polarization in the production channels $gb\to tH^-$ and  $g\bar{b}\to \bar{t}H^+$. We calculate the helicity amplitudes of the charged Higgs boson production and decay.Our calculation shows that the top quark from the charged Higgs boson decay provides a good probe for measuring $\tan\beta$, especially for the intermediate $\tan\beta$ region.  On the contrary, the top quark produced in association with the charged Higgs boson cannot be used to measure $\tan\beta$ because its polarization is highly contaminated by the $t$-channel kinematics.

\end{abstract}

\maketitle

\section{Introduction\label{Introduction}}

Recently, the ATLAS and CMS collaborations at Large Hadron Collider (LHC) have discovered a new boson with a mass of about 125 GeV. The detailed properties of the new particle are compatible with the Higgs boson in the standard model (SM).
One may expect that the latest discovery is just the beginning of the pursuit of new physics (NP) beyond the SM (BSM).
If BSM does exist, it is not strange that it is still hidden in the scalar sector.
As one of the simplest BSM models, the two-Higgs-doublet model (THDM) has been extensively investigated in the literature (for example, see Ref. \cite{Branco:2011iw} for
the latest review).
In the model five physical scalars emerge after spontaneous symmetry breaking, namely, the neutral CP-even Higgs boson ($h^0$ and $H^0$), the neutral CP-odd Higgs boson ($A^0$),~\footnote{If the separate discrete symmetry of two Higgs fields is not conserved, namely,in the CP-spontaneously broken THDM (Model IV in literature), three neutral scalars are not CP eigenstates; for example, see Refs. \cite{Huang:1999xa,Huang:2001me}.} and the charged Higgs boson ($H^{\pm}$). The $H^{\pm}$ boson is an undoubted NP signature.

In this work we focus on the charged Higgs boson production in the type-II THDM. In the model one Higgs doublet couples to up-type fermions while the other doublet couples to down-type fermions. After symmetry breaking the charged Higgs boson interacts with up- and down-type fermions as follows:
\beq
g_{H^-\bar{d}u}=\displaystyle\frac{g}{\sqrt{2}m_W} (m_d\tan\beta P_L+m_u\cot\beta P_R ), \label{coupling}
\eeq
where $\tan\beta$ is the ratio of the vacuum expectation values of the two-Higgs doublet and $P_{L/R}=(1\mp \gamma_5)/2$ is the chirality projector.
The chirality structure reflects in the polarizations of the quarks from the $H^\pm$ decay, which could be used to determine the value of $\tan\beta$ in the type-II THDM. Third generation quarks play an important role in measuring charged Higgs boson coupling because in the THDM such couplings are not severely suppressed by fermion mass, contrary to the light fermion case. Thanks to its heavy mass, the top quark decays promptly through the weak interaction such that all its quantum information is well kept in its decay products. Through the precision measurement of the decay products of the top quark, one can reconstruct the top quark event and measure the top quark polarization. One can then use the top quark polarization to measure $\tan\beta$~\cite{Huitu:2010ad,Godbole:2011vw,Gong:2012ru,Baglio:2011ap,Eriksson:2007fx}.

The charged Higgs boson can be directly produced in three channels: (1) $q \bar{q} \to \gamma^*/Z^* \to H^+ H^-$; (2) $g b \to t H^-~(g\bar{b}\to \bar{t} H^+)$; and (3) $q\bar{q}^\prime \to W^* \to AH^\pm/hH^\pm/HH^\pm$. As a heavy charged Higgs boson is preferred to satisfy the $b\to s\gamma$ constraint~\cite{Deschamps:2009rh}, the production rate of the $H^+H^-$ pair decreases dramatically with $m_{H^\pm}$. The $AH^\pm$ production rate depends on the unknown mass of the CP-odd Higgs boson $A$. In this work we focus our attention on the $t H^-/\bar{t}H^+$
associated production \cite{Barnett:1987jw,Olness:1987ep,Huang:1998vu,Borzumati:1999th,Zhu:2001nt,Plehn:2002vy}
\beq
g b \to t H^- \to t \bar{t} b, \qquad g \bar{b} \to \bar{t} H^+ \to \bar{t}t\bar{b}.
\eeq

In order to probe the  $H^+ \bar t b$ coupling using top polarization, one must be aware of two sources of the top production: (1) the top as the decay product of a charged Higgs boson; (2) the top produced in association with a charged Higgs boson. We demonstrate that the top or antitop quark produced in association with the charged Higgs boson is much less polarized than those from the $H^\pm$ decay.
For simplification we demand the charged Higgs boson entirely decays into a pair of top- and bottom-quarks.

This paper is organized as follows: In Sec.~\ref{heli-amp}, we calculate the helicity amplitudes of $H^- \to \bar{t} b$ and $g b \to t H^- $ processes, and obtain their degrees of top polarization respectively. In Sec.~\ref{collide-sim}, we simulate the $g b \to t H^- (\to \bar{t} b)$ process and its background events at LHC, and we plot the dependence of the degree of top polarization on $\tan\beta$ for the top quark from the $H^-$ decay. In Sec.~\ref{conclusion}, we conclude that the top quark from the charged Higgs boson decay provides a good probe for measuring the $\tan\beta$ while the top quark produced in association with the charged Higgs boson cannot be
used to measure $\tan\beta$.

\section{Helicity amplitude~\label{heli-amp}}
\subsection{Helicity amplitudes of $H^- \to \bar{t} b$}

\begin{figure}[b]
\begin{center}
\includegraphics[width=0.95\textwidth]{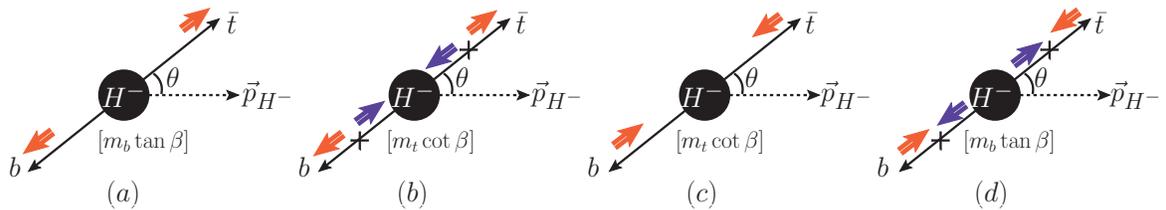}
\end{center}
\caption{\it Pictorial illustration of the helicity amplitude of $\bar{t}$ and $b$ from $H^-$ decay in the rest frame of $H^-$. The dashed-line arrows show the motion direction of $H^-$ in the c.m. frame. The long thin arrows display the $\bar t$ and $b$ moving directions, the short bold arrows along the long thin arrows denote the spin direction, and the cross symbols on the long thin arrows represent mass insertions which flip the fermion chirality. The Yukawa couplings are shown inside the square bracket.}
\label{hdecay}
\end{figure}

The matrix element of $H^- \to \bar{t} b$ is
\beq
\mathcal{M}_{\rm dec}(H^- \to \bar{t} b) = \frac{i g}{\sqrt{2} m_W} \overline{u}_b (m_b \tan\beta P_L + m_t \cot\beta P_R) v_{\bar{t}},
\label{mat_hdecay}
\eeq
yielding the helicity amplitude $\mathcal{M}(\lambda_b, \lambda_{\bar t})$ as follows:
\bea
\mathcal{M}_{\rm dec}(+,+) &=& \frac{i g}{\sqrt{2} m_W} \left( m_b\tan\beta \sqrt{(E_b + p)(E_t + p)} -m_t\cot\beta \sqrt{(E_b - p)(E_t - p)}  \right),\nonumber \\
\mathcal{M}_{\rm dec}(-,-) &=& \frac{i g}{\sqrt{2} m_W} \left(m_b\tan\beta \sqrt{(E_b - p)(E_t - p)} - m_t\cot\beta \sqrt{(E_b + p)(E_t + p)} \right) ,\nonumber \\
\mathcal{M}_{\rm dec}(+,-) &=& \mathcal{M}_{\rm dec}(-,+) = 0.
\label{hel_hdecay}
\eea
Clearly, the dependence of the helicity amplitudes on the polar angle $\theta_{\bar{t}H^-}$ is absent owing to the scalar feature of $H^\pm$. The $\theta_{\bar{t}H^-}$ is defined as the open angle between the $\bar{t}$ quark and the motion direction of $H^-$ in the c.m. frame; see Fig.~\ref{hdecay}. In order to conserve the spinless feature of the scalar boson, the two quarks from $H^-$ decay must exhibit the same helicity; see Eq.~\eqref{hel_hdecay} and Fig.~\ref{hdecay}.  Figure~\ref{hdecay} displays the helicity configuration of $b_R \bar{t}_R$, where (a) originates from the first term in Eq.~\eqref{mat_hdecay} while (b) comes from the second term in Eq.~\eqref{mat_hdecay} after double mass insertions.  The mass insertions lead to a suppression factor of $m_b m_t / m_{H^\pm}^2$, which is negligible for a heavy $H^\pm$.

To quantify the top quark polarization in the decay of $H^- \to \bar{t} b$, one introduces the degree of top quark polarization ($D$) defined as
\beq
D_{\rm decay}\equiv\frac{\Gamma(\bar{t}_L) - \Gamma(\bar{t}_R)}{\Gamma(\bar{t}_L) + \Gamma(\bar{t}_R)} = \frac{(m_t \cot\beta)^2 - (m_b\tan\beta)^2}{(m_t\cot\beta)^2 + (m_b\tan\beta)^2},
\eeq
where we ignore the double mass insertion terms.

\begin{figure}
\begin{center}
\includegraphics[width=0.4\textwidth]{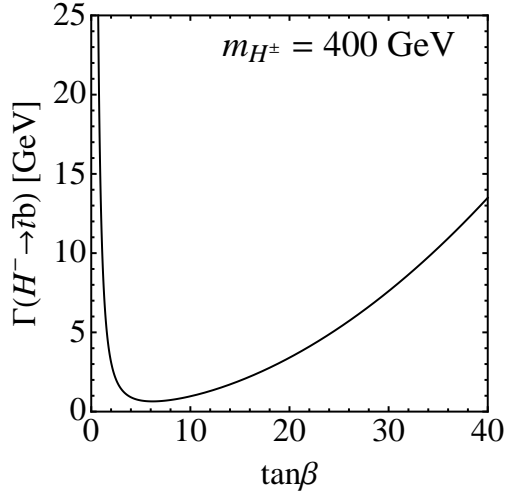}
\end{center}
\caption{\it Partial decay width of $H^- \to \bar{t} b$ as a function of $\tan\beta$ for $m_{H^\pm}=400~{\rm GeV}$, with $m_b^{\overline{MS}}=4.18$~GeV and
$m_t^{\overline{MS}} =160.0$~GeV in the coupling of Eq. \ref{coupling}.
}
\label{hdecay_width}
\end{figure}

The decay width of $H^- \to \bar{t} b$ is
\beq
\Gamma(H^- \to \bar{t} b) =\frac{g^2 N_c m_{H^\pm}}{32\pi m_W^2}\left(m_b^2\tan^2\beta + m_t^2 \cot^2\beta \right) \left(1-\frac{m_t^2}{m_{H^\pm}^2}\right)^2,
\eeq
where $N_c=3$ denotes the color factor. In the above equation we ignore the bottom-quark mass except for those in the Yukawa couplings. The decay width, as shown in Fig.~\ref{hdecay_width}, is highly sensitive to $\tan\beta$.

\subsection{Helicity amplitudes of $g b \to t H^- $}

The matrix elements of the process of $gb \to t H^- $ are
\bea
i\mathcal{M}_{s} &=& \left(t^a\right) \frac{g g_s}{\sqrt{2}m_W}\frac{1}{\hat{s}} ~\bar{u}_t \left(m_b\tan\beta P_R + m_t \cot\beta P_L\right) (\not{\! \ell} + m_b)\not{\! \epsilon}u_b~, \nonumber \\
i\mathcal{M}_{t} &=& \left(t^a\right)\frac{g g_s}{\sqrt{2}m_W}\frac{1}{\hat{t}-m_t^2}~\bar{u}_t \not{\! \epsilon}  (\not{\! q} + m_t) \left(m_b\tan\beta P_R + m_t \cot\beta P_L\right) u_b~
\eea
for $s$- and $t$-channel processes respectively (see Feynman diagrams in Fig.~{\ref{fig:signal_feyndiag-gb-tH}).
Here $\ell = p_g + p_b$, $q = p_t - p_g $, $\hat s= \ell^2$, $\hat t= q^2$ and $\epsilon$ represents the polarization vector of the incoming gluon. Denote the helicity amplitudes as
$$\mathcal{M}_{\rm prod}(\lambda_g, \lambda_b,\lambda_t),$$
where $\lambda_{i} = + (-)$ labels the right-handed (left-handed) helicity of the particle $i$ in the overall c.m. frame.

\begin{figure}
\begin{center}
\includegraphics[width=0.7\textwidth]{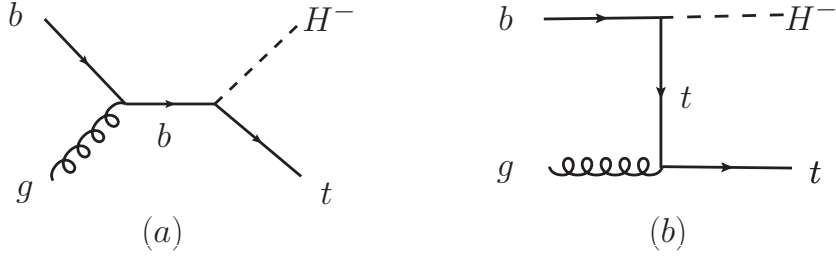}
\end{center}
\caption{\it The Feynman diagrams of the process $b g \to t H^-$: (a) $s$ channel, (b) $t$ channel.}
\label{fig:signal_feyndiag-gb-tH}
\end{figure}

The momenta of the incoming and outgoing patrons are chosen as follows:
\begin{align}
&p_g= (\sqrt{\shat}/2, 0, 0, \sqrt{\shat}/2), ~~~\qquad p_b = (\sqrt{\shat}/2, 0, 0, -\sqrt{\shat}/2), \nonumber \\
&p_t = (E_t, p\sin\theta, 0, p\cos\theta), \qquad
p_{H^-} = (E_{H^-}, -p\sin\theta, 0, -p\cos\theta).
\end{align}
Apart from a common factor of $(gg_s t^a)$, nonzero helicity amplitudes of the $s$-channel diagram are
\bea
\mathcal{M}_s(+,+,+) &=& -\frac{m_b\tan\beta}{m_W}\frac{\sqrt{E_t -p}}{\shat^{1/4}} \cos\frac{\theta}{2},
\label{sppp} \\
\mathcal{M}_s(+,+,-) &=& ~~\frac{m_b\tan\beta}{m_W}\frac{\sqrt{E_t +p}}{\shat^{1/4}} \sin\frac{\theta}{2},
\label{sppm} \\
\mathcal{M}_s(-,-,+) &=& \frac{m_t\cot\beta}{m_W}\frac{\sqrt{E_t +p}}{\shat^{1/4}} \sin\frac{\theta}{2},
\label{smmp} \\
\mathcal{M}_s(-,-,-) &=& \frac{m_t\cot\beta}{m_W}\frac{\sqrt{E_t -p}}{\shat^{1/4}} \cos\frac{\theta}{2}.\label{smmm}
\eea
The incoming bottom quark and gluon exhibit the same helicity in order to produce a spin-$1/2$  fermion in the $s$-channel propagator. The chirality of the top quark has to be opposite to the chirality of the bottom quark, owing to the Yukawa coupling. Hence, if the top quark and bottom quark have the same helicity, then there must be a mass insertion on the external top quark fermion line. It yields a weight factor of $\sqrt{E_t -p}$ which vanishes in the limit of $m_t \to 0$; see Eqs.~\eqref{sppp} and \eqref{sppm}.  In Fig.~\ref{prod_s_hel} we show a pictorial demonstration of the helicity configurations.  It is also easy to verify that the spatial angle distributions are
\bea
\mathcal{M}_s(+,+,+) \sim \mathcal{M}_s(-,-,-) ~\propto ~~d_{1/2,1/2}^{1/2}(\theta)=\cos\frac{\theta}{2},\nonumber \\
\mathcal{M}_s(+,+,-) \sim \mathcal{M}_s(-,+,+) ~\propto ~~d_{1/2,-1/2}^{1/2}(\theta)=\sin\frac{\theta}{2}.\nonumber
\eea

\begin{figure}
\begin{center}
\includegraphics[width=0.95\textwidth]{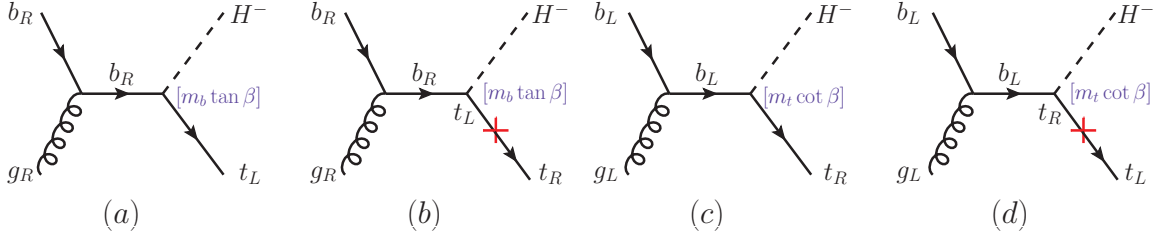}
\end{center}
\caption{\it Pictorial illustration of the helicity amplitude of the process of $gb\to t H^- $ in the overall c.m. frame. The terms inside square brackets represent the interactions. The cross symbols on the long thin arrows represent mass insertions which flip the fermion chirality.}
\label{prod_s_hel}
\end{figure}

The $t$-channel diagram is more complicated as it involves a higher orbit angular momentum. It has eight nonzero helicity amplitudes, which are given as follows:
\bea
\mathcal{M}_t(+,+,+) &=& ~~\frac{m_b \tan\beta}{m_W} \frac{1}{\shat^{1/4}}\left(\frac{\sqrt{E_t -p}(E_{H^-}-p\cos\theta)}{E_t - p \cos\theta} +\frac{m_t\sqrt{E_t+p}}{E_t - p \cos\theta} \right)\cos\frac{\theta}{2}, \label{tppp} \\
\mathcal{M}_t(+,+,-) &=& -\frac{m_b \tan\beta}{m_W} \frac{1}{\shat^{1/4}}\left(\frac{\sqrt{E_t +p}(E_{H^-}-p\cos\theta)}{E_t - p \cos\theta} +\frac{m_t\sqrt{E_t-p}}{E_t - p \cos\theta} \right)\sin\frac{\theta}{2}, \label{tppm} \\
\mathcal{M}_t(+,-,+) &=&-\frac{m_t\cot\beta}{m_W}\frac{\sqrt{E_t +p}}{\shat^{1/4}}\frac{p\sin\theta}{E_t - p \cos\theta}\cos\frac{\theta}{2},\label{tpmp} \\
\mathcal{M}_t(-,+,+) &=&~~-\frac{m_b\tan\beta}{m_W}\frac{\sqrt{E_t -p}}{\shat^{1/4}}\frac{p\sin\theta}{E_t - p \cos\theta}\sin\frac{\theta}{2},\label{tmpp} \\
\mathcal{M}_t(+,-,-) &=&~~\frac{m_t\cot\beta}{m_W}\frac{\sqrt{E_t -p}}{\shat^{1/4}}\frac{p\sin\theta}{E_t - p \cos\theta}\sin\frac{\theta}{2},\label{tpmm} \\
\mathcal{M}_t(-,+,-) &=&~~-\frac{m_b\tan\beta}{m_W}\frac{\sqrt{E_t +p}}{\shat^{1/4}}\frac{p\sin\theta}{E_t - p \cos\theta}\cos\frac{\theta}{2},\label{tmpm} \\
\mathcal{M}_t(-,-,+) &=& ~~-\frac{m_t \cot\beta}{m_W} \frac{1}{\shat^{1/4}}\left(\frac{\sqrt{E_t +p}(E_{H^-}-p\cos\theta)}{E_t - p \cos\theta} +\frac{m_t\sqrt{E_t-p}}{E_t - p \cos\theta} \right)\sin\frac{\theta}{2}, \label{tmmp} \\
\mathcal{M}_t(-,-,-) &=& ~~-\frac{m_t \cot\beta}{m_W} \frac{1}{\shat^{1/4}}\left(\frac{\sqrt{E_t -p}(E_{H^-}-p\cos\theta)}{E_t - p \cos\theta} +\frac{m_t\sqrt{E_t+p}}{E_t - p \cos\theta} \right)\cos\frac{\theta}{2}.\label{tmmm}
\eea
We explicitly single out the $s$ channel-like contributions in Eqs.~\eqref{tppp},~\eqref{tppm},~\eqref{tmmp},~\eqref{tmmm}. Note the sign difference between the $s$ channel and $t$ channel, which clearly implies a destructive interference between the $s$ channel and $t$ channel.

The degree of top polarization of the $gb\to t H^- $ process is
\bea
D_{\rm prod}(\shat) &\equiv& \frac{\hat{\sigma}(t_R)-\hat{\sigma}(t_L)}{\hat{\sigma}(t_R)+\hat{\sigma}(t_L)} =\frac{\displaystyle\int d\Phi_2 ~\left[\mathcal{A}(t_R) - \mathcal{A}(t_L)\right]}{\displaystyle \int d\Phi_2 ~\left[\mathcal{A}(t_R) + \mathcal{A}(t_L)\right] },
\eea
where $\hat{\sigma}$ denotes the cross section of the hard scattering process in the c.m. frame of the $gb$ system, and
\beq
\mathcal{A}(t_R) \equiv  \sum_{\lambda_g,\lambda_b}\left|\mathcal{M}_{\rm prod}(\lambda_g,\lambda_b, +)\right|^2,\qquad
\mathcal{A}(t_L) \equiv  \sum_{\lambda_g,\lambda_b}\left|\mathcal{M}_{\rm prod}(\lambda_g,\lambda_b, -)\right|^2.
\eeq
It is straightforward to show that the $H^-$-$t$-$b$ couplings can be factorized out in $D_{\rm prod}$ as follows:
\bea
D_{\rm prod}(\shat) &=& \frac{(m_t\cot\beta)^2-(m_b \tan\beta)^2}{(m_t\cot\beta)^2+ (m_b \tan\beta)^2 } \times \hat{R}_{\rm prod}.
\label{dilution_def}
\eea
The first term parameterizes the top quark polarization generated solely by the Yukawa coupling of $H^-$-$t$-$b$, which gives rise to the maximal degree of polarization of the top quark produced with $H^-$. Due to the higher partial waves in the $t$-channel process, the top quark polarization is diluted by the top quark's angular momentum. The dilution factor $\hat{R}_{\rm prod}$ depends only on the top quark's kinematics.
Figure~\ref{dilution} shows $\hat{R}_{\rm prod}$ as a function of the c.m. energy $\sqrt{\hat{s}}$ of the hard scattering process for $m_{H^\pm}=400~{\rm GeV}$ (solid black) and $m_{H^\pm}=600~{\rm GeV}$ (dashed red). The magnitude of $\hat{R}_{\rm prod}$ is less than 0.5, which suppresses the degree of top quark polarization; that is why we call it the dilution factor.
Furthermore, one should note that the dilution factor is negative in the threshold region of the $t H^- $ pair and turns positive in the large invariant mass region. The sign of the dilution factor is important because it is the key to determining the top quark polarization in the charged lepton angle distribution. After convoluting with the parton distribution functions, the dilution factor might ruin our whole analysis of the top quark polarization.
The paper~\cite{Huitu:2010ad} shows that at the hadron level in the $t H^- $ production the degree of polarization decreases when the mass of a charged Higgs boson increases; the degree of polarization is smaller at 14~TeV than at 7~TeV for LHC collision.
Therefore, we focus our attention on measuring top quark polarization in the $H^\pm$ decay rather than the $t H^- $ production in this work.

\begin{figure}
\begin{center}
\includegraphics[width=0.4\textwidth]{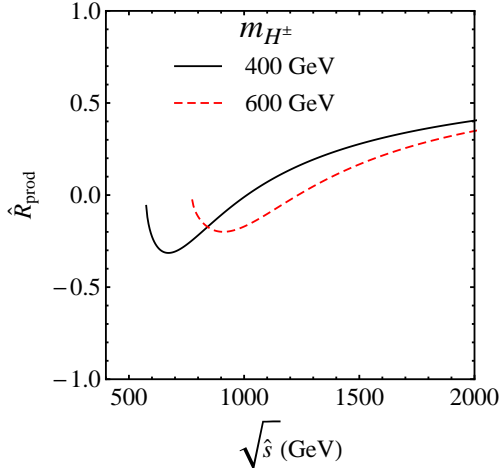}
\end{center}
\caption{\it The dilution factor in Eq.~\eqref{dilution_def} as a function of the energy of the overall c.m. frame ($\sqrt{\shat}$) with $m_{H^\pm}=400$GeV (solid black) and $m_{H^\pm}=600$GeV (dashed red). }
\label{dilution}
\end{figure}

\section{Collider simulation\label{collide-sim}}

In this section we preform a detailed collider simulation of the $t H^-  / \bar{t} H^+$ pair production and the dominant SM backgrounds. Figure \ref{fig:rate} shows the cross section of the $gb \to t H^- $ scattering as a function of $\tan\beta$. After exploring the potential of observing the signal events at the 14 TeV LHC with an integrated luminosity of $100~{\rm fb}^{-1}$, we examine the top quark polarization measurements and comment on the possibility of determining $\tan\beta$ in the long run.
\begin{figure}
\begin{center}
\includegraphics[width=0.4\textwidth]{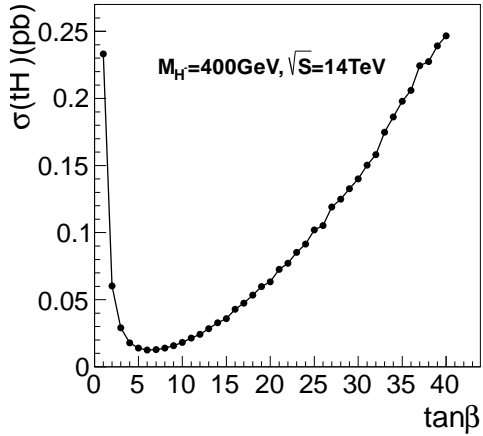}
\end{center}
\caption{\it Inclusive production cross section of the $gb \to t H^- $ scattering as a function of $\tan\beta$ at the 14~TeV LHC for $m_{H^\pm}=400~{\rm GeV}$, $m_b^{\overline{MS}}=4.18~{\rm GeV}$, and $m_t^{\overline{MS}} =160.0~{\rm GeV}$, in the coupling of Eq. \ref{coupling}. }
\label{fig:rate}
\end{figure}

Since we are interested in the polarization of the top quark or antitop quark from the charged Higgs boson decay,  both $gb \to t H^-  $ and $g\bar{b} \to \bar{t} H^+$ processes contribute to the signal events as they both lead to $t\bar{t}j_b$ ($j_b$ denotes the $b$-tagged jet which originates from either a $b$ or $\bar{b}$ quark). In reality one cannot tell a $b$-jet and a $\bar{b}$ jet apart. In order to measure the top quark polarization, we demand $\bar{t} \to \ell^- \bar{b} \bar{\nu}$ and use the angle distribution of $\ell^-$ to measure the top quark polarization.  We further require $t \to b j j $ to increase the signal rate.~\footnote{The leptonic decay $t\to \ell^+ b \nu$ can also be used to determine the top quark polarization. The two missing neutrinos can be fully reconstructed from the missing energy and the on-shell conditions of the $W$ bosons and $t$ quarks~\cite{Berger:2010fy,Zhang:2010kr}.}
It then yields an event topology of $j_b j_b j_b j j \ell^- $ plus missing energy as follows:
\begin{align}
&g\bar{b}\rightarrow \bar{t} H^+\rightarrow (W^-\bar{b}) (t \bar{b})\rightarrow (\ell^-\bar{\nu_\ell}\bar{b})(j j b \bar{b}), \nonumber \\
& gb \rightarrow t H^-\rightarrow (W^+ b) (\bar{t} b) \rightarrow (j j b)(\ell^- \bar{\nu_\ell} \bar{b} b).
\end{align}

\begin{figure}
\begin{center}
\includegraphics[width=0.7\textwidth]{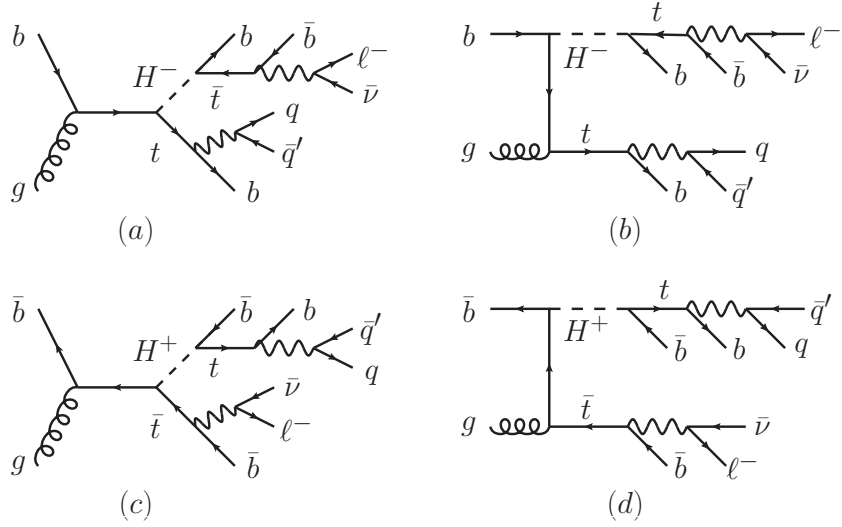}
\end{center}
\caption{\it The Feynman diagrams of  the process $pp \to t H^- \to t\bar{t} b$ (a,~b) and the process $pp\to \bar{t} H^+ \to t \bar{t} \bar{b}$ (c,~d), where the subsequent decays of $t \to jjb $ and
$\bar{t}\to \ell^- \bar{\nu_\ell} \bar{b}$ are considered. }
\label{fig:signal_feyndiag}
\end{figure}

Figure~\ref{fig:signal_feyndiag} displays the Feynman diagrams of both signal processes with subsequent decays.  At the LHC the $gb$ and $g\bar{b}$ initial states  give rise to the same production rate.
 Only half of $\bar{t}$ quarks in the signal event sample are from the $H^-$ decay, and their polarization is completely determined by the chirality structure of the $g_{H^{-} \bar{t} b}$ coupling.  The other half of $\bar{t}$ quarks emerge at the production level, whose polarization is highly diluted. In order to reveal the connection between the top quark polarization and $\tan\beta$, one has to find a set of optimal cuts to separate the $t H^- $ and $\bar{t} H^+$ signal events.

Two SM backgrounds are considered in this work:
\bea
t\bar{t}b &:& pp \to t\bar{t} j_b\to b W^+ \bar{b} W^- j_b \to j_b j_b j_b jj \ell^- \bar{\nu},\nonumber\\
t\bar{t}j     &:& pp \to t\bar{t}j \to b W^+ \bar{b}W^- j \to j_b j_b jjj \ell^- \bar{\nu}.
\eea
The $t\bar{t}b$ background is irreducible as it contributes exactly the same event topology as the signal. On the other hand,
the reducible {\bf $t\bar{t}j$} background could mimic the signal when the light jet $j$ (denoting the light-flavor quarks or gluons) is misidentified as a $b$ jet.

\subsection{Event selection}

In this work we adapt MadGraph/MadEvent~\cite{Alwall:2011uj} to generate both the signal and background processes.
The following basic cuts are applied while generating events in MadGraph5,
\begin{align}
&p_T^j \geq 10~{\rm GeV}, \qquad~|\eta_{j}| \leq 5.0~, \nonumber \\
&p_T^{\ell^-}>10~{\rm GeV},\qquad |\eta_{\ell^-}| \leq 2.5~, \nonumber \\
&\!\!\!\not{\!\rm E}_T> 20~{\rm GeV},\qquad ~\Delta R_{jj,j\ell}>0.4~,
\label{basic-cuts}
\end{align}
where $p_T$ denotes the transverse momentum, $\not{\!\rm E}_T$ is the missing transverse momentum from the invisible neutrino in the final state, and $\Delta R$ is the separation in the azimuthal angle ($\phi$) pseudorapidity ($\eta$) plane between the objects $a$ and $b$
\beq
\Delta R_{ab} \equiv \sqrt{\left(\eta_a -\eta_b\right)^2 + \left(\phi_a - \phi_b \right)^2}.
\eeq

Table~\ref{tbl-efficiency} displays the number of events for the signals and backgrounds at the 14 TeV LHC with an integrated luminosity of $100~{\rm fb}^{-1}$ for $m_{H^\pm}=400~{\rm GeV}$ and three values of $\tan\beta$. The dominant SM background is from the $t\bar{t}j$ production which is about 3 or 4 orders of magnitude larger than the signal; see the third row in the table.  A series of kinematic cuts is needed to extract the small signal out of the tremendous backgrounds. We optimize the cut criteria specially for $\tan\beta=6$ where the signal is difficult to detect.

\begin{table}
\begin{center}
\caption{Number of events of the signal and backgrounds at the 14 TeV LHC with an integrated luminosity of $100~{\rm fb}^{-1}$ for $m_{H^\pm}=400~{\rm GeV}$ and three values of $\tan\beta$. }
\label{tbl-efficiency}
\begin{tabular}{c||c|c|c|c|c|c||c|c}
\hline
$\tan\beta$ & \multicolumn{2}{c|}{1} & \multicolumn{2}{c|}{6} & \multicolumn{2}{c||}{40} & \multicolumn{2}{c}{SM backgrounds}\tabularnewline
\hline
 & $tH^{-}$ & $\bar{t}H^{+}$ & $tH^{-}$ & $\bar{t}H^{+}$ & $tH^{-}$ & $\bar{t}H^{+}$ & $t\bar{t}j$ & $t\bar{t}b$\tabularnewline
\hline
\hline
Inclusive rate & 23310 & 23300 & 1255 & 1227 & 24660 & 23520 & $1.075\times10^{7}$ & 234000\tabularnewline
Hard $p_T$ cuts & 11843 & 13466 & 687 & 719 & 14421 & 13890 & $2.12\times10^{6}$ & 25052\tabularnewline
$\Delta M_{\bar{t}j_{\rm extra}}$ & 4980 & 368 & 672 & 20 & 5680 & 383 & 39238 & 386\tabularnewline
$p_{T}(j_{{\rm extra}})$ & 3910 & 305 & 532 & 16 & 4375 & 310 & 14942 & 171\tabularnewline
$b$ tagging & 2346 & 183 & 312 & 10& 2625 & 186 & 299 & 102\tabularnewline
\hline
\hline
Number of events & \multicolumn{2}{c|}{2529} & \multicolumn{2}{c|}{322} & \multicolumn{2}{c||}{2811} & \multicolumn{2}{c}{401}\tabularnewline
$S/B$ & \multicolumn{2}{c|}{6.3} & \multicolumn{2}{c|}{0.8} & \multicolumn{2}{c||}{7.0} & \multicolumn{2}{c}{$-$}\tabularnewline
$S/\sqrt{B}$ & \multicolumn{2}{c|}{126.3} & \multicolumn{2}{c|}{16.1} & \multicolumn{2}{c||}{140.3} & \multicolumn{2}{c}{$-$}\tabularnewline
$\sqrt{S+B}$ & \multicolumn{2}{c|}{54.1} & \multicolumn{2}{c|}{26.9} & \multicolumn{2}{c||}{56.7} & \multicolumn{2}{c}{$-$}\tabularnewline
\hline
\end{tabular}
\end{center}
\end{table}

\subsection{Hard $p_T$ cut on jets}

Our signal events consist of five jets in the final sate. As they originate from a heavy scalar decay, the top quark and bottom quark are boosted such that they have a hard $p_T$. As a result, the jets from the energetic top quark are also highly boosted to yield a hard $p_T$. On the contrary, the top quarks in the SM backgrounds are predominantly produced in the threshold region where the top quarks are not highly boosted. The jets from top quark decays in the backgrounds tend to be soft.  We examine the $p_T$'s of the final state jets and impose hard cuts on their $p_T$'s to suppress the SM backgrounds.

We order the five jets in the final state by their $p_T$ values in each event:
\beq
p_T(j_{\rm 1st}) > p_T(j_{\rm 2nd}) > p_T(j_{\rm 3rd}) > p_T(j_{\rm 4th}) > p_T(j_{\rm 5th}).
\eeq
\begin{figure}
\begin{center}
\includegraphics[width=0.3\textwidth,clip]{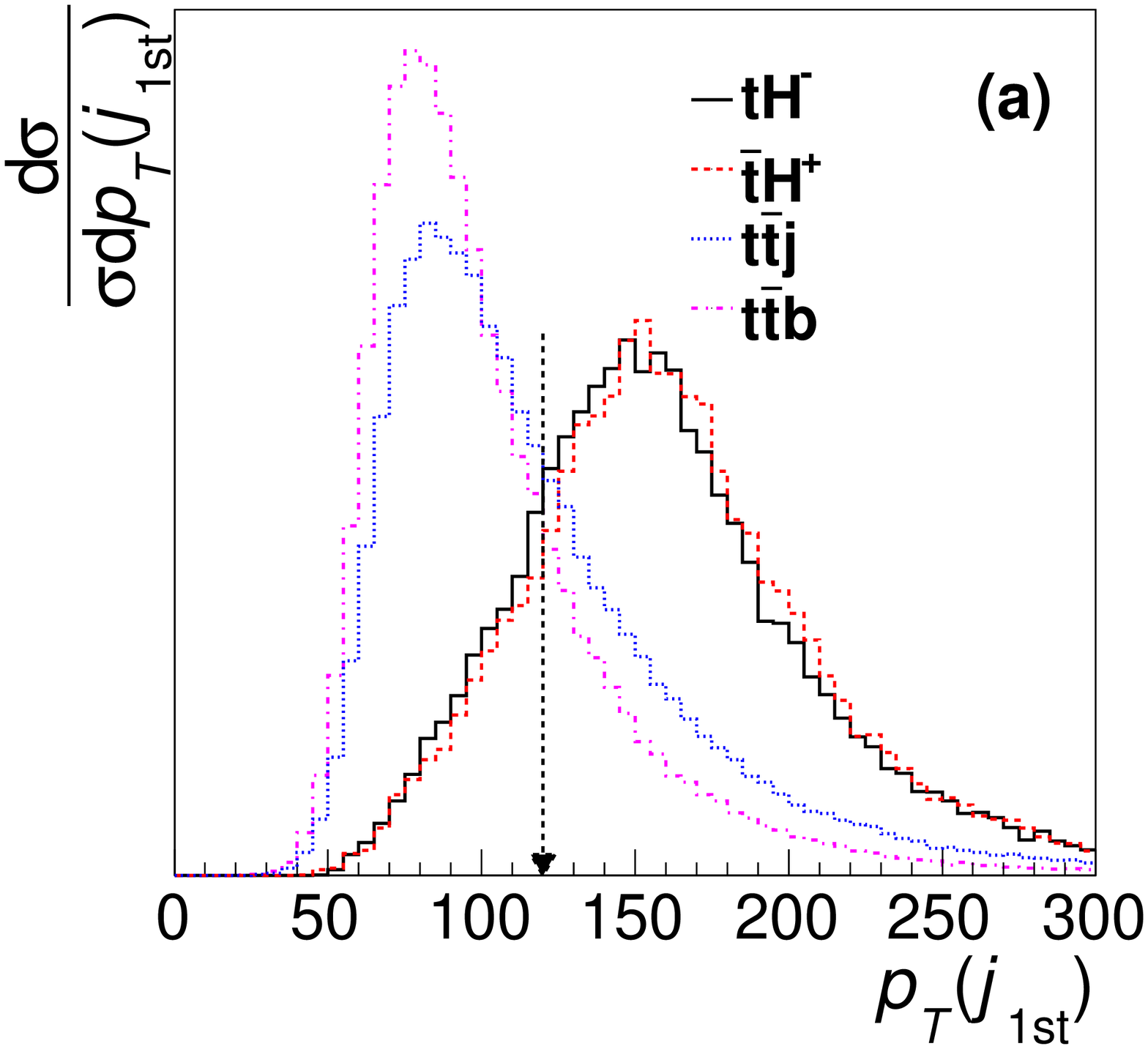}
\includegraphics[width=0.3\textwidth,clip]{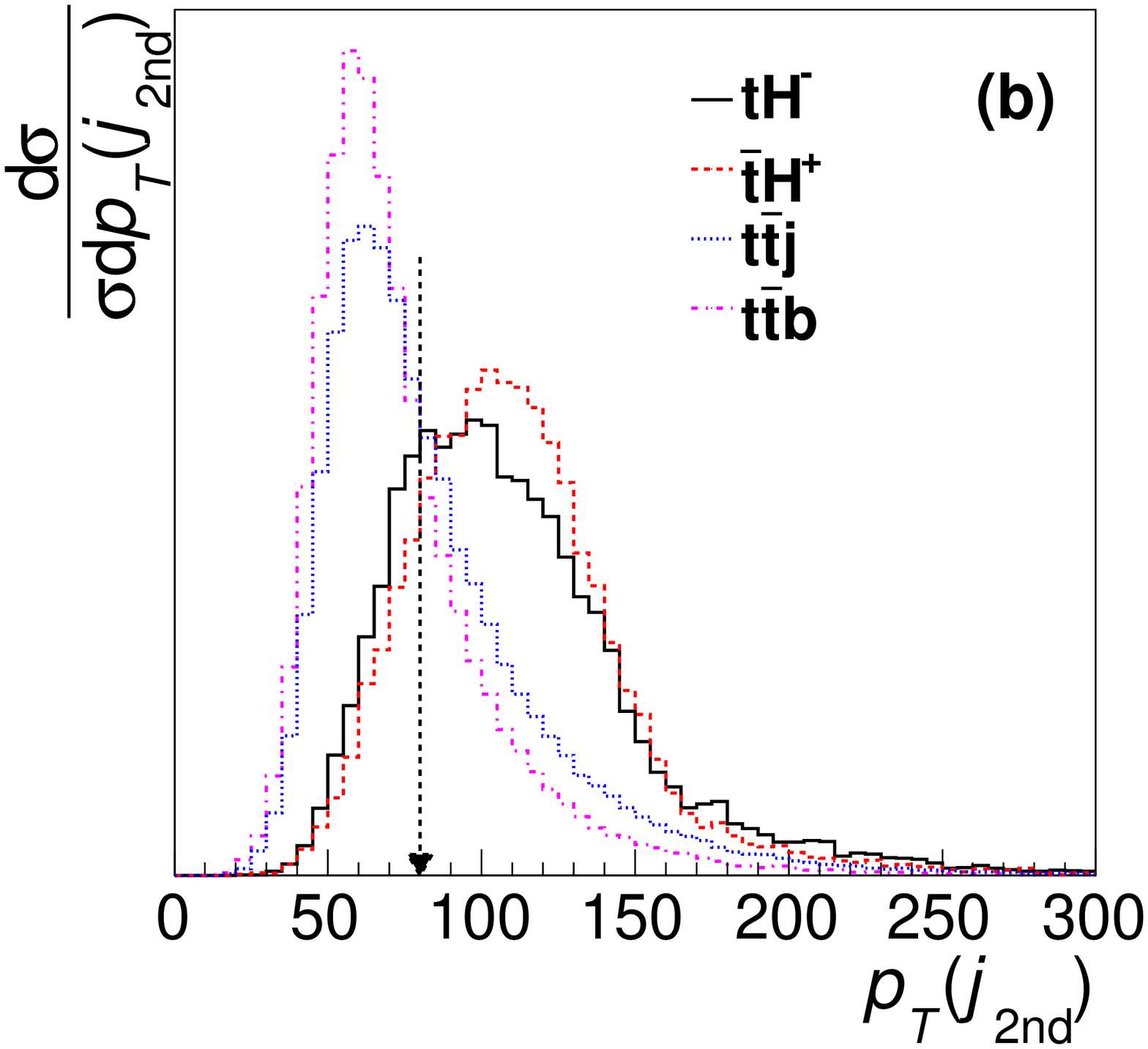}
\includegraphics[width=0.3\textwidth,clip]{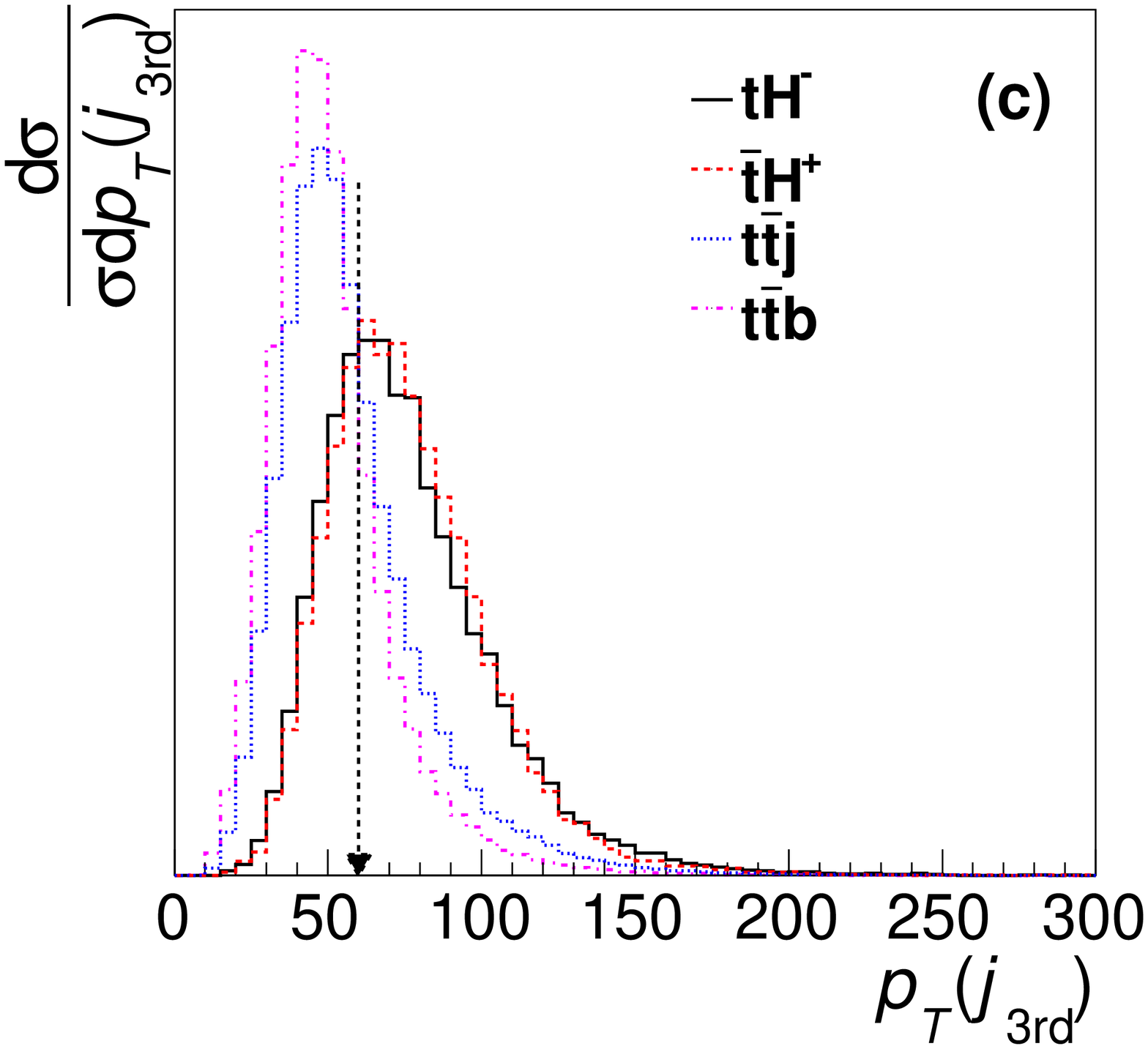}
\end{center}
\caption{\it Normalized $p_T$ distributions of the first, second and third jets ordered by their $p_T$ values in each event for $\tan\beta=6$. The vertical dashed-line arrows show the hard $p_T$ cut imposed. }
\label{ptjets}
\end{figure}
Figure~\ref{ptjets} displays the normalized $p_T$ distributions of the $p_T$ ordered jets: (a) the leading (first) jet; (b) the second jet;  (c) the third jet.  It clearly shows that the signal events exhibit much harder $p_T$ distributions than the SM background events.
It enables us to impose hard $p_T$ cuts on the first three hard jets to suppress the SM backgrounds.
In this study we adapt a  set of hard $p_T$ cuts as follows:
\beq
p_T(j_{\rm 1st}) \geq 120~{\rm GeV}, \qquad p_T(j_{\rm 2nd})\geq 80~{\rm GeV}, \qquad p_T(j_{\rm 3rd})>60~{\rm GeV};
\label{hardcut}
\eeq
see the vertical dashed-line arrows in Fig.~\ref{ptjets}. For example, the leading $p_T$ jet is very often the $b$ or $\bar{b}$ jet from the charged Higgs boson decay such that its $p_T$ distribution peaks around $(m_{H^\pm}^2 - m_t^2)/(2m_{H^\pm})\simeq 160~{\rm GeV}$; see the solid black and dashed red curves in Fig.~\ref{ptjets}(a). On the other hand, the leading $p_T$ jet in the background is often the $b$ jet from top quark decay and naturally exhibits a peak around $m_t/3 \sim 60~{\rm GeV}$ in the $p_T$ distribution; see the dotted blue and dashed-dotted magenta curves.

The fourth row in Table~\ref{tbl-efficiency} shows the number of events after the hard cuts for both signals and backgrounds. The hard cuts remove almost half of the signal events, but they suppress the backgrounds by almost an order of magnitude.  At this stage of analysis, the $t\bar{t}j$ and $t\bar{t}b$ backgrounds still dominate over the signal processes after the hard $p_T$ cuts given in Eq.~\eqref{hardcut}. The backgrounds are about 80 times larger than the signals for small and large $\tan\beta$ ($\tan\beta=1$ or $\tan\beta=40$). For $\tan\beta =6$ the ratio of the background and signal is about 1500.

\subsection{Top reconstruction, mass-window cut and extra-jet tagging}

So far we treat all the jets equally and no $b$ tagging is performed. The key to suppress the $t\bar{t}j$ background is to identify the extra jet (denoted as $j_{\rm extra}$, the jet produced in association with the $t\bar{t}$ pair) as a $b$ jet. The minimal $\chi^2$-template method~\cite{Berger:2011xk} is adapted to reconstruct the $t\bar{t}$ pair and singles out the extra jet.

The minimal $\chi^2$-template method is based on the $W$-boson and top quark masses to select the extra jet. For each event we loop over all jet combinations and pick the combination which minimizes the following $\chi^2$:
\beq
\chi^2 = \frac{(m_W - m_{jj})^2}{\Delta m_W^2} + \frac{(m_t - m_{j\ell^- \bar{\nu}})^2}{\Delta m_t^2} + \frac{(m_t - m_{jjj})^2}{\Delta m_t^2}.
\eeq
Note that the invisible neutrino needs to be fully determined in order to reconstruct $\bar{t}$ in above equation. The transverse momentum of the neutrino can be determined from the momentum imbalance in the transverse plane while the longitudinal momentum of the neutrino ($p_{\nu L}$) can be derived from the $W$-boson on-mass-shell condition, $ m_{\ell^- \bar{\nu}}^2 = \left( p_{\ell^-} + p_{\bar \nu}\right)^2= m_W^2$.
It yields a twofold solution as
\beq
p_{\bar{\nu} L}=\frac{1}{2p_{\ell^- T}^2}\left[\left(m_W^2+2\overset{\rightarrow }{P}_{\ell^-T}\cdot\!\!\overset{\rightarrow}{\not{\!\rm E}}_T\right)p_{\ell^-L}\pm E_{\ell^-}\sqrt{\left(m_W^2+2\overset{\rightarrow }{P}_{\ell^{-}T}\cdot\!\!\overset{\rightarrow }{\not{\!\rm E}}_T\right)^2-4p_{\ell^-T}^2 \!\!\not{\!\rm E}_T^2}~\right],
\eeq
when
$\left(m_W^2+2\overset{\rightarrow }{P}_{\ell^{-}T}\cdot\!\!\overset{\rightarrow }{\not{\!\rm E}}_T\right)^2-4p_{\ell^-T}^2 \!\!\not{\!\rm E}_T^2 \geq 0 $. The ambiguity of the twofold solution is removed also by the minimal $\chi^2$ requirement. The method is very efficient at identifying the extra jet and reconstructing the leptonically decayed top quark, but it is less efficient at reconstructing the hadronically decayed top quark because of the combinatorial ambiguities of jets in the final state~\cite{Berger:2011xk}.

\begin{figure}
\begin{center}
\includegraphics[width=0.5\textwidth,clip]{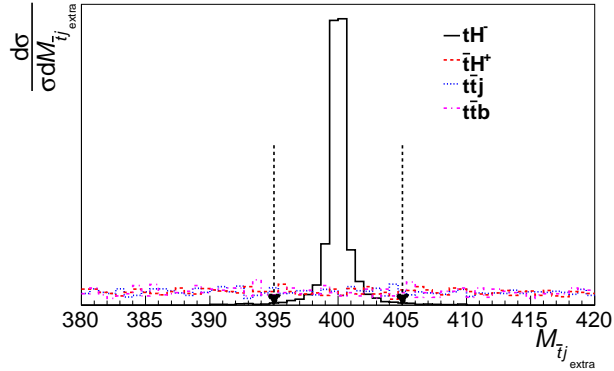}
\end{center}
\caption{\it Normalized invariant mass distribution of the reconstructed $\bar{t}$ and $j_{extra}$ pair for $\tan\beta=6$. The vertical dashed lines show the mass-window cut. }
\label{fig:tbmass}
\end{figure}
With a reconstructed $t$ quark and $\bar{t}$ quark in hand, we are ready to reconstruct $H^-$. Figure~\ref{fig:tbmass} shows the normalized invariant mass distribution of the $\bar{t}$ and $j_{\rm extra}$ pair, denoted as $m_{\bar{t}j_{\rm extra}}$. Since our signal events consist of both $t H^- $ and $\bar{t} H^+$, one half of the signal events have a sharp peak in the $m_{\bar{t}j_{\rm extra}}$ distribution (solid black curve) when the $\bar{t}$ and $j_{\rm extra}$ pair indeed comes from the $H^-$ decay. The pin shape reflects the narrow width of the charged Higgs boson.  On the contrary, the other half of the signal events exhibit a fairly broad bump (dashed red curve) as $\bar{t}$ is not from the $H^-$ decay in the $g\bar{b} \to \bar{t} H^+$ process. Similarly, the two SM backgrounds also show a broad spectrum and peak in the low mass region. To further improve the signal-to-background ratio, we require the invariant mass of the reconstructed $\bar{t}$ quark and the extra jet to be within the mass window,
\beq
\Delta M_{\bar{t}j_{\rm extra}}\equiv \left| M_{\bar{t} j_{\rm extra}}-M(H^\pm) \right| \leq 5{\rm GeV}, \label{cut-m-tmj}
\eeq
where $5~{\rm GeV}$ is the expected experimental resolution for a 400~GeV $H^-$ decaying into a $\bar{t} b$ pair.
As shown in Table~\ref{tbl-efficiency}, both the $t\bar{t}j$ and $t\bar{t}b$ backgrounds are reduced by a factor of around 60 at the cost of the reduction rate in signal by $20\% \sim 50\%$. That increases the signal-to-background ratio by a factor of $12$. The reduction of the signal is mainly from the $\bar{t}H^+$ process which exhibits a continuous nonresonance spectrum of the $\bar{t}j_{\rm extra}$ invariant mass; see the dashed red curve in Fig.~\ref{fig:tbmass}.  Such a nonresonance feature is simply because the $\bar{t}$ and the extra jet are not from the $H^-$ decay. Note that, when $\tan\beta=6$,  the mass-window cut does not have much impact on the $tH^-$ signal process because the $H^-$ boson width is very narrow such that almost all the signal events fall inside the mass window.

\begin{figure}
\begin{center}
\includegraphics[width=0.5\textwidth,clip]{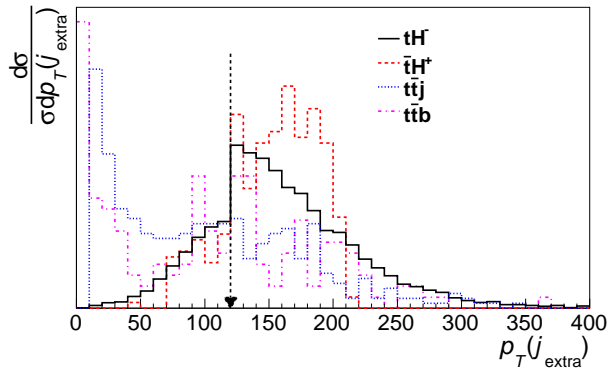}
\end{center}
\caption{\it Normalized $p_T$ distribution of the extra jet for $\tan\beta=6$.}
\label{fig:ptjexta}
\end{figure}

The extra jet in the signal often originates from the heavy $H^-$ decay and tends to have a large $p_T$. The extra jet in the SM background, predominately from QCD radiation,  tends to have a much softer $p_T$.  In Fig.~\ref{fig:ptjexta} we plot the normalized $p_T$ distribution of the extra jet. It is clear that both $t H^- $ (solid black) and $\bar{t} H^+$ (dashed red) signal processes have a large $p_T$ while the background processes peak in the small $p_T$ but still have a long tail in the large $p_T$ region. We impose the following cut on $p_T(j_{\rm extra})$ to optimize the signal:
\beq
p_T(j_{\rm extra}) \geq 120~{\rm GeV}.
\eeq
The cut suppresses the background rate by a factor of 3 and keeps about $2/3$ of the signal events; see the sixth row in Table~\ref{tbl-efficiency}.

Finally, we demand that the extra jet be a $b$ jet and choose the tagging efficiency as $60\%$ and a moderate mistagging efficiency as $2\%$~\cite{Aad:2009wy}. The very last requirement sufficiently eliminates the SM backgrounds; see the seventh row in Table~\ref{tbl-efficiency}.  At the bottom of the table we also show the number of both signal and background events after applying all the cuts, the signal-to-background ratio, the statistical significance of the signal, and the statistical uncertainties in the measured signal event rate.
It is very promising to observe the signal events after the optimal cuts. The significance of the signal is well above $5\sigma$ for a broad range of $\tan\beta$. For $\tan\beta=6$, more than 300 signal events survive, which could be used to probe top quark polarization.

\section{top quark polarization and $\tan\beta$ measurement}

Armed  with the reconstructed $\bar{t}$, $j_{\rm extra}$ and $H^-$, we are ready to measure $\bar{t}$-quark polarization.
In the helicity basis, the $\bar{t}$-quark polarization can be found from the distribution in $\theta_{\rm hel}$, the angle of the charged lepton in the rest frame of $\bar{t}$ quark relative to the top quark direction of motion in the rest frame of $H^-$. The angular correlation of $\ell^-$ is given by \beq
\frac{d\Gamma}{\Gamma d\cos\theta_{\rm hel}}=\frac{1}{2} (1\pm\cos\theta_{\rm hel}),
\label{lepton-ang}
\eeq
 where the ($+$) choice is for a left-handed $\bar{t}$ quark while ($-$) is for a right-handed $\bar{t}$ quark.
The angular distribution of $\ell^+$ in the rest frame of a $t$ quark is the same
but with the ($+$) choice for the right-handed $t$ quark while ($-$) is for the left-handed $t$ quark.
Note that the above formula is
insensitive to the  next-to-leading order (NLO) QCD corrections, which are only at $O(10^{-3})$ \cite{Czarnecki:1990pe}. The tree-level analysis based on top polarization will not be altered too much even including NLO QCD corrections.

In Fig.~\ref{costheta}(a), we plot the normalized distribution of $\cos\theta_{\rm hel}$ of the signal and background processes with basic cuts in Eq.~\eqref{basic-cuts}. For the signal processes we choose $m_{H^\pm}=400~{\rm GeV}$ and $\tan\beta=1$. The $tH^-$ signal events exhibit a clear shape of $(1+\cos\theta_{\rm hel})/2$.  On the contrary, the distribution of the $\bar{t}H^+$ signal is flat, owing to the effects of higher partial waves generated by the $t$-channel diagram. Both the $t\bar{t}j$ and $t\bar{t}b$ backgrounds also show a flat distribution. The top quark polarization possessed by the $tH^-$ signal is diluted by the $\bar{t}H^+$ signal and the other two SM backgrounds.

In order to demonstrate the dependence of top quark polarization on $\tan\beta$ in the $tH^-$ signal events, we plot the normalized distribution of $\cos\theta_{\rm hel}$  of the $tH^-$ channel for $m_{H^\pm}=400~{\rm GeV}$ and three values of $\tan\beta$ in Fig.~\ref{costheta}(b): $\tan\beta=1$ (solid), $\tan\beta=6$ (dashed), and $\tan\beta=40$ (dotted). The top quark is highly polarized for $\tan\beta=1$ and $\tan\beta=40$ but is nearly unpolarized for $\tan\beta=6$. The results remain the same even after all cuts.

\begin{figure}
 \begin{center}
  \includegraphics[width=0.4\textwidth]{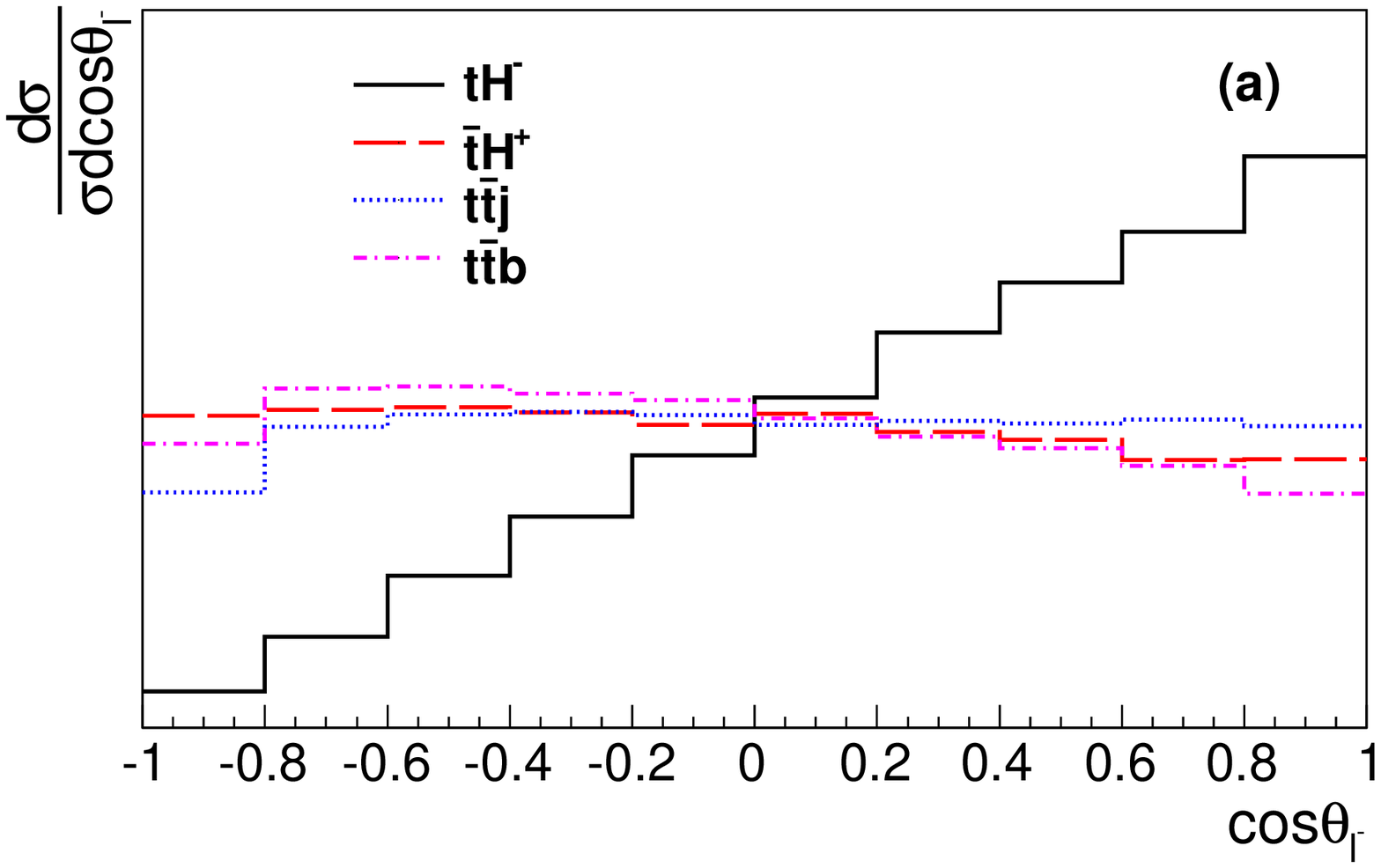}
  \includegraphics[width=0.4\textwidth]{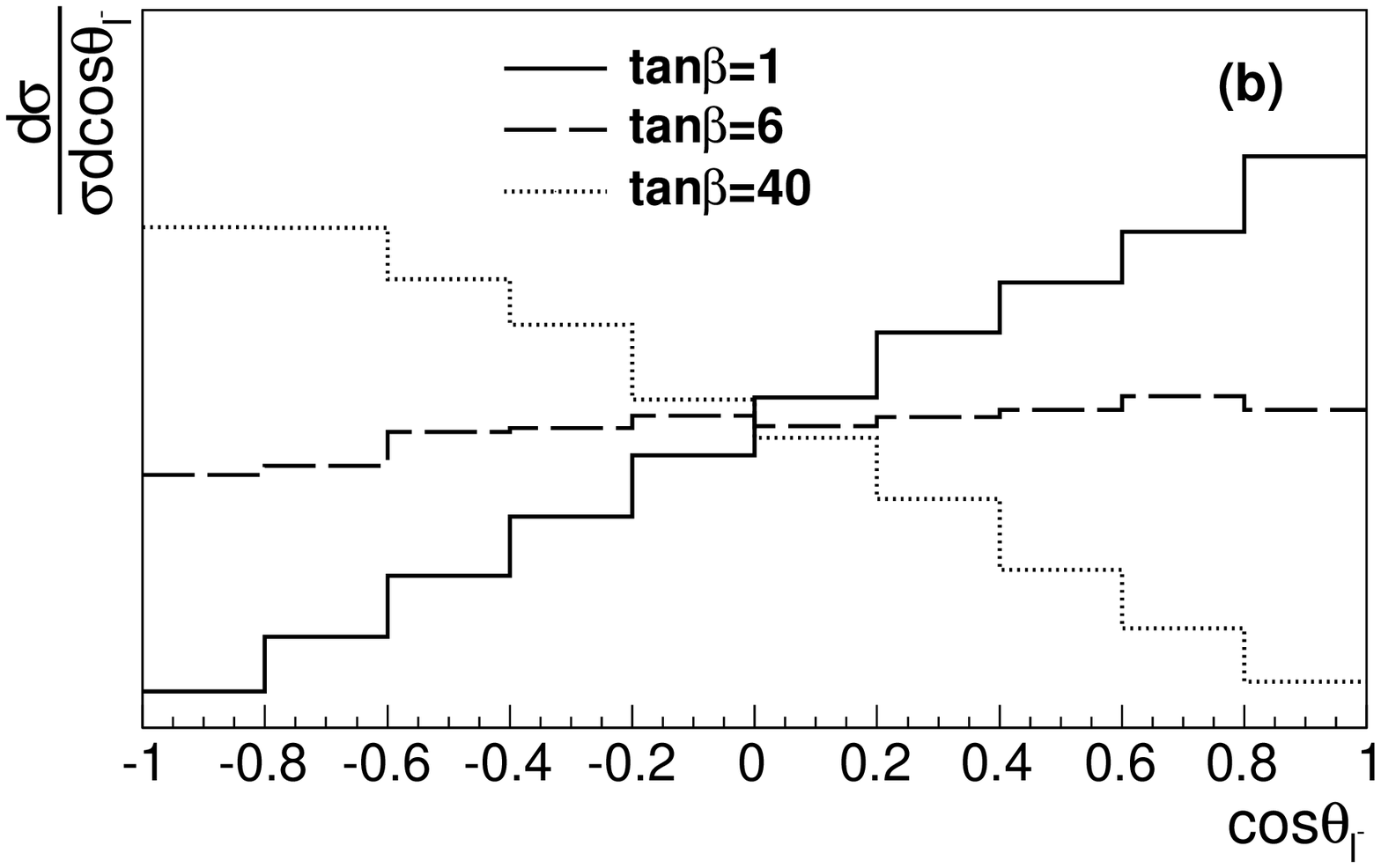}
  \end{center}
  \caption{\it (a) Normalized distribution of $\cos\theta_{\rm hel}$ of the signal and background processes with basic cuts in Eq.~\eqref{basic-cuts} at the 14 TeV LHC with an integrated luminosity of $100~{\rm fb}^{-1}$ for $m_{H^\pm}=400~{\rm GeV}$ and $\tan\beta=1$: $tH^-$ (solid black), $\bar{t}H^+$ (dashed red), $t\bar{t}j$ (dotted blue) and $t\bar{t}b$ (dotted-dashed magenta). (b) Normalized distribution of $\cos\theta_{\rm hel}$ of the $tH^-$ signal events for three benchmark values of $\tan\beta$: $\tan\beta=1$ (solid), $\tan\beta=6$ (dashed), and $\tan\beta=40$ (dotted).}
  \label{costheta}
\end{figure}

The degree of top quark polarization can be written as the ratio
\beq
D=\frac{N_+ - N_-}{N_+ + N_-},
\eeq
where $N_+$ ($N_-$) is the number of right-handed (left-handed) polarized top quarks in the helicity basis;
for the antitop quark case $N_+$ ($N_-$) is the number of left-handed (right-handed) polarized antitop quarks.
Correspondingly the angle distribution of $\theta_{\rm hel}$  could be written as
\beq
\frac{d\Gamma}{\Gamma d\cos\theta_{\rm hel}}=\frac{1}{2} (1+D\cos\theta_{\rm hel}).
\eeq
Simple algebra leads to the following identity:
\beq
D=3 \int^1_{-1} \cos\theta_{\rm hel} \frac{d\Gamma}{\Gamma d\cos\theta_{\rm hel}} d\cos\theta_{\rm hel}.
\label{d-definition}
\eeq
We obtain the degree of top quark polarization from the $\cos\theta$ distribution, which is divided into 10 bins:
\beq
D=3\sum_{i=1}^{10} \cos\theta_i \left(\frac{d\sigma}{\sigma d\cos\theta}\right)_i\Delta \cos\theta
 =\frac{3\sum_{i=1}^{10} \cos\theta_i N_i}{\sum_{i=1}^{10} N_i},
\eeq
where $\cos\theta_i$ is the middle point value of each bin, $(\frac{d\sigma}{\sigma d\cos\theta})_i$ is the normalized value for each bin, $\Delta \cos\theta$ is the bin width and $N_i$ is the number of events falling into each bin. We consider the statistical error for $N_i$ as $\Delta N_i=\sqrt{N_i}$; the statistical error for the degree of top quark polarization is calculated as
\beq
\Delta D=\sqrt{\sum_{i=1}^n\left|\frac{\partial D}{\partial N_i}\right|^2 \left({\Delta N_i}\right)^2}.
\label{d-error}
\eeq

Based on the degree of polarization we can easily get the spin fraction $F_{\pm}$,
\beq
F_{\pm}\equiv\frac{N_{\pm}}{N_- + N_+}= \frac{1\pm D}{2},
\eeq
the fraction of top quarks with spin along the basis direction.
We can also define the asymmetry $A_{FB}$ of the distribution of $\cos\theta_{\rm hel}$ as
\beq
A_{FB}\equiv\frac{\sigma_F -\sigma_B}{\sigma_F + \sigma_B},
\eeq
where
\beq
\sigma_F \equiv \int^1_0 \frac{d\sigma}{\sigma d\cos\theta_{\rm hel}} d\cos\theta_{\rm hel},\quad
\sigma_B \equiv \int^0_{-1} \frac{d\sigma}{\sigma d\cos\theta_{\rm hel}} d\cos\theta_{\rm hel}.
\eeq
It is easy to check that without imposing any kinematic cut $D=2A_{FB}$, but the relation would break down after the kinematic cut. Table~\ref{top} displays the degree of polarization $D$, polarization fraction $F_+$, and asymmetry $A_{FB}$ before cuts and after selection cuts and event reconstruction. From the table we can see that for $\tan\beta=1, 40$ the top quark is highly polarized, the kinematic cuts change the degree of polarization by $10\%$; the relation of $D=2A_{FB}$ still holds.
 \begin{table}
\begin{center}
\caption{Degree of polarization $D_{\rm decay}$, polarization fraction $F_+$ and asymmetry $2A_{FB}$ before cuts and after cuts and event reconstruction for signal and background processes respectively. }
\label{top}
\begin{tabular}{|c|c|c|c|c|c|c|c|}
\hline
\multirow{2}[2]*{$\tan\beta$}& & \multicolumn{2}{c|}{$D_{\rm decay}$} & \multicolumn{2}{c|}{$F_+$} & \multicolumn{2}{c|}{$2A_{FB}$}
\\\cline{2-8}
 && No cut & After cut & No cut & After cut & No cut & After cut\tabularnewline
\hline
 \multirow{2}[2]*{1}&$tH^{-}$ & 0.97 & 0.87 &0.99  & 0.93 & 0.98 & 0.88\\\cline{2-8}
&$\bar{t}H^{+}$& $-0.10$ & $-0.43$ & 0.45 & 0.29 & $-0.09$ & $-0.41$ \tabularnewline
\hline
\multirow{2}[2]*{6}&$tH^{-}$ & 0.12 & $-0.08$ & 0.56 &0.46  &0.10&$-0.08$\\\cline{2-8}
&$\bar{t}H^{+}$& $-0.00$ & $-0.02$ &0.50 &0.49 &0.00 &$-0.03$\tabularnewline
\hline
\multirow{2}[2]*{40}&$tH^{-}$ & $-0.89$ & $-1.03$ & 0.06 &$-0.02$  &$-0.90$&$-1.03$\\\cline{2-8}
&$\bar{t}H^{+}$& 0.10 & $-0.03$ &0.55 &0.49 &0.09 &$-0.02$\tabularnewline
\hline
&$t\bar{t}j$ &0.06  & 0.09 & 0.53 & 0.55 &0.03  & 0.04\tabularnewline
\hline
&$t\bar{t}b$ & $-0.14$ &0.03  &0.43  &0.51  & $-0.17$ & 0.25\tabularnewline
\hline
\end{tabular}
\end{center}
\end{table}

Figure~\ref{pt-sb}(a) shows the degree of polarization of the antitop quark as a function of $\tan\beta$ in the $tH^-$ signal process for $m_{H^\pm}=400~{\rm GeV}$. The $2A_{FB}$ is also plotted for comparison. The top quark polarization is a good probe for a wide range of $\tan\beta$.
The intermediate $\tan\beta$ has been considered very hard to measure. Figure~\ref{pt-sb} shows that the $D_{\rm decay}$ varies rapidly in the region of $\tan\beta=5\sim10$. This feature enables us to determine $\tan\beta$ using top polarization. However, the degree of polarization cannot be used to determine the value of $\tan\beta$ in the large $\tan\beta$ region as the degree of polarization approaches -1. Including the $\bar{t}H^+$ signal and the two SM backgrounds inevitably reduces the degree of polarization, as depicted in Fig.~\ref{pt-sb}(b). The green band [cf. Eq.~\eqref{d-error}] shows the statistical uncertainties derived from all the signal and background events after all the kinematic cuts and event reconstructions.

\begin{figure}
 \begin{center}
  \includegraphics[width=0.4\textwidth]{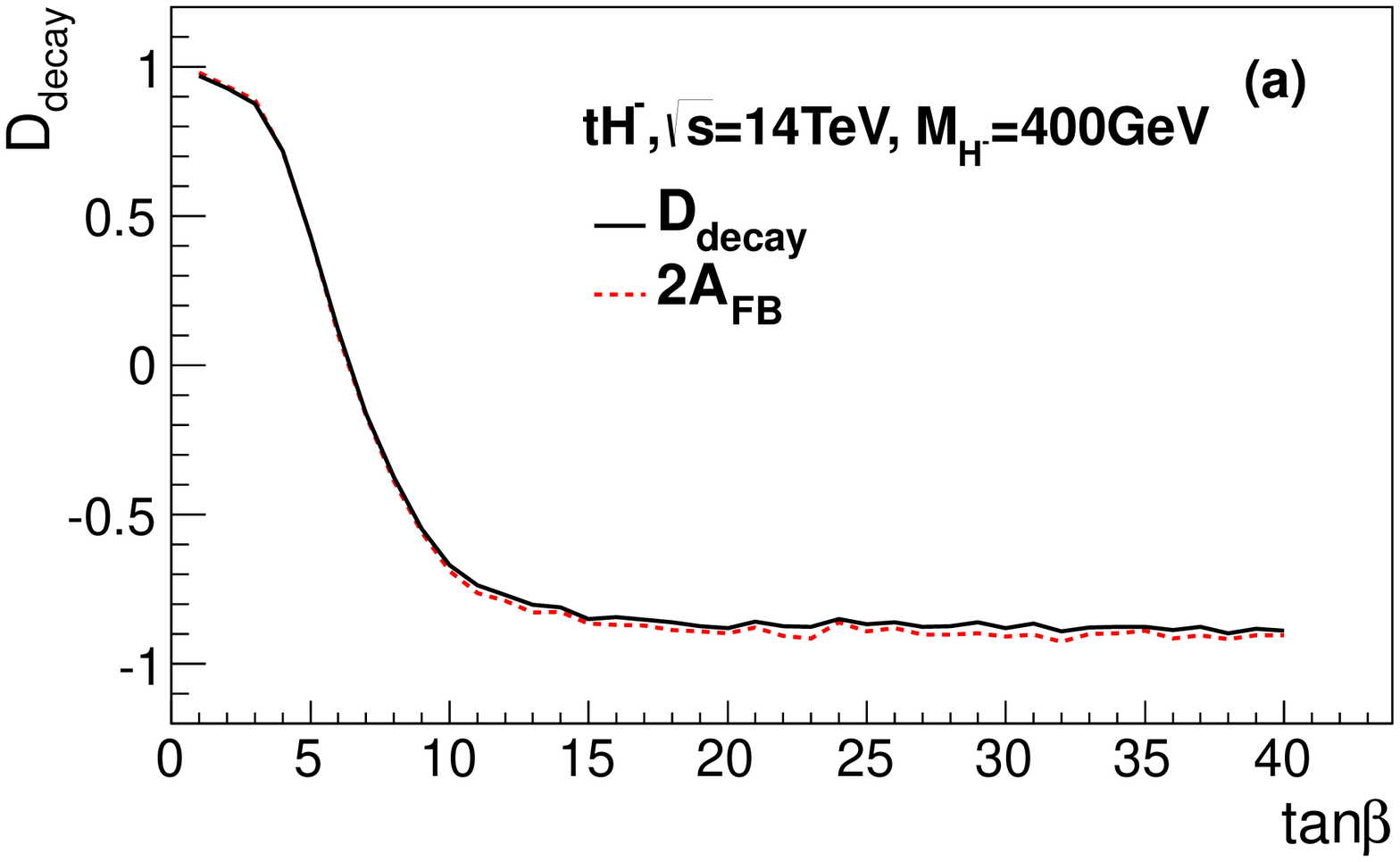}
  \includegraphics[width=0.4\textwidth]{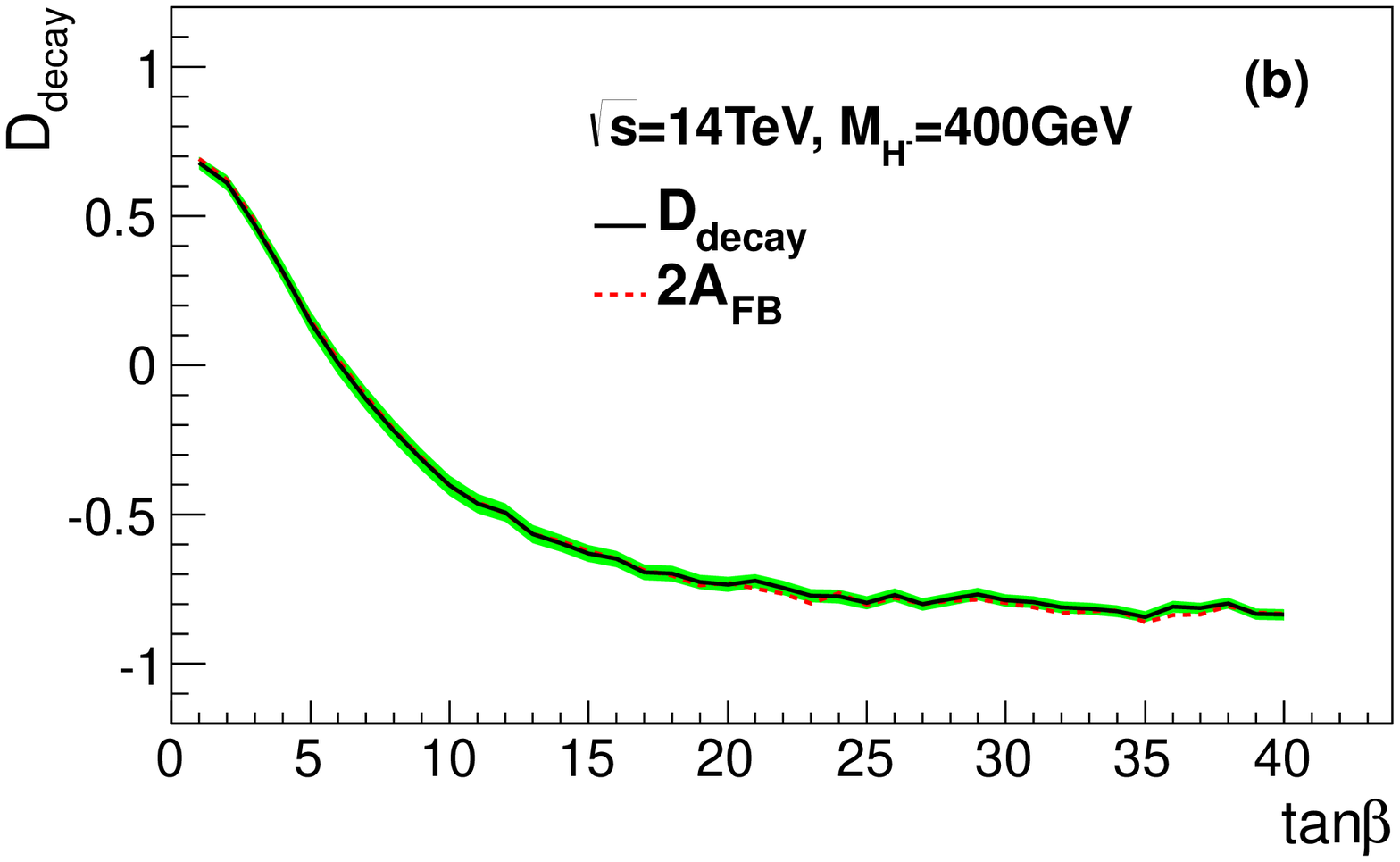}
  \end{center}
  \caption{\it (a) The degree of polarization of the antitop quark as a function of $\tan\beta$ of the $tH^-$ signal event and (b) of all the signal and background processes with $m_{H^\pm}=400~\rm{GeV}$. The solid black curve shows the degree of polarization defined in Eq.~\eqref{d-definition}; the dashed red curve shows $2A_{FB}$. The green band in (b) represents only the statistical uncertainties.}
  \label{pt-sb}
\end{figure}

\section{Conclusion and Discussion\label{conclusion}}
The charged Higgs boson, an undoubted signal of new physics, appears in many new physics models. In the type-II two-Higgs-doublet model the chirality structure of the coupling of charged Higgs boson to the top and bottom quarks is very sensitive to the value of $\tan\beta$. As the polarization of the top quark can be measured experimentally from the top quark decay products, one could make use of the top quark polarization to determine the value of $\tan\beta$. In this work we preform a detailed analysis of measuring top quark polarization in the charged Higgs boson production channels $gb\to tH^-$ and  $g\bar{b}\to \bar{t}H^+$. We calculate the helicity amplitudes of the charged Higgs boson production and decay. Our calculation shows that the top quark from the charged Higgs boson decay provides a good probe for measuring $\tan\beta$, especially for the intermediate $\tan\beta$ region.  On the contrary, the top quark produced in association with the charged Higgs boson cannot be used to measure $\tan\beta$ because its polarization is highly contaminated by the $t$-channel kinematics.

The analysis in this paper is based on tree-level estimation for signal and background, and we would like to comment on the higher-order effects. The NLO QCD corrections to the $gb \to H^- t$ process have been calculated in Ref.~\cite{Zhu:2001nt}. It is shown that the ratio of the NLO cross section to the LO cross section varies roughly from $1.6$ to $1.8$ when the charged Higgs boson mass increases from 200 to 1000 GeV. The NLO QCD corrections can reduce the scale dependence of the LO cross section.
In order to simulate the real collider environment, one needs to use a full parton shower,
 including both the initial state radiation (ISR) and the final state radiation (FSR), to calculate more precisely the physical observables, e.g., the jet multiplicity, jet transverse momentum and energy, etc.
Our signal events consist of five jets from heavy resonance decays such that the jets exhibit a large
transverse momentum; see Fig.~\ref{ptjets}. On the other hand, the ISR and FSR tend to produce
soft jets, which are not often able to pass the stringent cuts imposed on the $p_T$ of jets in our analysis.
Our results should not vary dramatically by the ISR and FSR effects. Needless to say,
it is necessary to perform a thorough analysis including the parton shower effects to
get a more realistic prediction on the signal discovery potential, top quark reconstruction
efficiency, the uncertainty of measuring top quark polarization, etc. But it is beyond
the scope of the current paper and should be presented elsewhere.

\begin{acknowledgments}
We thank Chen Zhang for checking helicity amplitudes and many useful discussions. Q.H.C. is supported in part by the National Science Foundation of China under Grant No. 11245003. The work of X.W., X.P.W. and S.H.Z. was supported in part by the Natural Science Foundation of China under Grants No. 11075003 and No. 11135003. X.W. is also supported in part by the China Postdoctoral Science Foundation under Grant No. 2012M520098.
\end{acknowledgments}

%\bibliographystyle{apsrev}
%\bibliography{reference}

\end{document}